# Strong and variable stratospheric CO emission from lava-planet 55 Cnc e observed with NIRCam/JWST

Ignas Snellen[1], Yamila Miguel[2,1], Leoni Janssen[1], Darío Gonzalez Picos[1], Sam de Regt[1], Natalie Grasser[1], Lars Klijn[1]

*[1] Leiden Observatory, Leiden University, Postbus 9513, 2300 RA, Leiden, The Netherlands*

*[2] SRON Netherlands Institute for Space Research, Niels Bohrweg 4, 2333 CA, Leiden, The Netherlands*

**Some rocky planets orbit so close to their host stars that stellar heating melts their surfaces[1,2]. They offer a rare glimpse of planets in a magma-ocean state, providing an observable analogue to processes that likely shaped the early Earth and other terrestrial planets during their infancy. Recent JWST observations[3,4] of five eclipses of the prototypical lava planet 55 Cnc e have confirmed earlier hints[5,6] that it exhibits highly variable thermal emission, with low-resolution spectroscopy pointing to a possible volatile-rich atmosphere likely rich in CO and $CO_2$. Here we report on an analysis of the same JWST datasets but at their native spectral resolution, utilizing cross-correlation techniques[7]. An unambiguously strong ~8σ signal from CO in emission is recovered during one out of five epochs, with potential ~3σ detections during two others. The strongest observed cross-correlation signal is difficult to reconcile with a hydrostatic atmosphere, requiring a steep and strong thermal inversion at the right pressure level (~1-10 mbar) and a relative abundance of $CO_2$ that is at least 3 orders of magnitude lower which would otherwise mask the CO signal. Self-consistent atmospheric modelling indicates that this is most readily achieved in a hydrogen-rich atmosphere, which produces the steepest inversions and highest $CO/CO_2$ ratios. The pronounced epoch-to-epoch variability suggests that the CO signal does not trace a static atmosphere alone, but may reveal a transient, dynamically active component, potentially linked to variable atmospheric outflow.**

Our analysis focuses on JWST secondary eclipse observations of 55 Cnc e (four from program ID-2084 and one from ID-1952) using the NIRCAM long-wave 3.9 – 5 μm spectral channel with the GRISMR element[8,9]. Technical details are provided in the original studies[3,4] and summarized in the Methods Section. Both preceding studies focus on eclipse spectra binned by 30 to 50 wavelength steps to decrease the noise and emphasize potential broad-band features. Here we analyse the data at its native resolving power of R = 1600 to 1700, sampled with wavelength steps of $9.8\times10^{-4}$ μm. This resolution enables the use of cross-correlation techniques widely used in ground-based exoplanet atmosphere studies[7] and recently pioneered[10,11] with JWST/Nirspec, which optimally combine potential signals from individual molecular lines to lower the detection threshold and enable unambiguous detections of atmospheric species. In the case of 55 Cnc e, cross-correlation analyses require special care due to the high planet orbital velocity[12] of 228 km $s^{-1}$. Exoplanet secondary eclipse spectroscopy is strictly a misnomer, because the planet is not visible during secondary eclipse but instead during the baseline observations before and after the eclipse. The five visits each cover an orbital phase range of 0.35 to 0.60 with a gradual change of the planet radial velocity from +190 km $s^{-1}$ to -140 km $s^{-1}$, meaning a shift of ~5 wavelength steps which needs to be compensated for to avoid blurring any signal.

The analysis started with fully calibrated exposure products (level 2c) downloaded from MAST (see Methods Section). For each exposure, a one-dimensional spectrum was extracted from the 2D detector frame using a simple extraction scheme without signal-to-noise weighting, including pixels that on average contributed more than 5% of the total flux at each wavelength. Each individual spectrum is subsequently interpolated to the planet rest frame, after which the baseline at each wavelength is fitted with a combination of an exponential decay and a linear function to normalize the light curve independently at each wavelength step. The planet-to-star flux ratio for each visit is subsequently $\langle S_{base}\rangle / \langle S_{ecl}\rangle - 1$, where $\langle S_{ecl}\rangle$ and $\langle S_{base}\rangle$ are the average spectra during eclipse (T2 – T3) and baseline (<T1 or >T4). The top panel of Figure 1 shows the colour-scale spectral light curves for the five individual visits and their average, binning each 100 spectra and by 5 pixels in spectral direction. The secondary eclipse is marginally visible during Visit 1 while clearly detected during the other visits. Horizontal blocks and striping indicate residual systematic effects at a level up to 100-200 ppm. This has been reported and extensively discussed in the preceding studies[3,4], but their origin remains unknown. It is less relevant for the cross-correlation analysis performed here since these systematic effects appear to be largely common-mode (i.e. achromatic) in nature. The lower panel in Figure 1 shows both the binned planet-to-star contrast spectra for each visit, and the average combined spectrum at native resolution. The contrast during Visit 1 (~50 ppm) and Visit 4 (~200 ppm) seem weaker and stronger compared to the other visits, although the latter may be influenced by what appears to be a significant systematic effect during the first half of the eclipse.

Cross-correlation template spectra were generated with petitRADTRANS[13] for hydrostatic atmospheres consistent with the mass and radius[14] of 55 Cnc e. We adopted inverted temperature–pressure profiles spanning $\left(P_{\rm high}\text{–}P_{\rm low},\ T_{\rm min}\text{–}T_{\rm max}\right)$, bounded by isothermal layers above and below. We focus on CO because its strong, well-separated lines make it far more detectable at high spectral resolution than in the binned spectra used previously: the cross-correlation sensitivity scales approximately as $\sqrt{R}$, implying a gain of roughly 5–7 over earlier studies. By contrast, the densely packed $CO_2$ lines merge into a broad feature with little structure at $R = 1500$, and are therefore not expected to yield a measurable cross-correlation signal. We considered pure CO atmospheres, CO mixed with $H_2$ as a spectrally inactive background gas, and models with additional $CO_2$, which can mask the CO lines.

The left panel of Figure 2 shows the cross-correlation signal for Visit 1, using a spectral template that assumes a pure CO atmosphere with a thermal inversion between pressures of $10^{-2}$ and $10^{-3}$ bar, spanning temperatures from 1000 to 3500 K, as illustrated in the left inset. In addition to the main cross-correlation peak, a series of repeating aliases is visible, arising from the semi-regular spacing of the CO lines, as is evident from the template autocorrelation function shown by the orange curve. The inset on the right presents the $K_p$–$V_{\rm sys}$ diagram[7], which maps the cross-correlation signal over a range of assumed planetary orbital velocities, $K_p$. The signal is strongest at $K_p \sim 200$ km s$^{-1}$, close to the expected value of 228 km s$^{-1}$, confirming that the CO signal originates from the planet rather than the star. The right panel of Figure 2 shows the phase-dependent trail of the CO signal, that is, the cross-correlation signal as a function of time in the stellar rest frame. Although the signal-to-noise is modest, the signal is visible over much of the observation, except during secondary eclipse when the planet is obscured by the star. Before eclipse, the CO signal is redshifted as the planet moves toward superior conjunction; after eclipse, it becomes blueshifted as the planet moves away again and the radial component of its orbital velocity points toward Earth.

To investigate the strength of the cross-correlation signal, we generated a broad grid of planetary spectra with petitRADTRANS, varying the pressure level of the thermal inversion. These spectra were converted to planet-to-star contrast ratios and combined with Gaussian noise matched to the observed planetary spectrum to produce mock cross-correlation functions, whose amplitudes were then compared with the observed signal-to-noise ratio. Because the observed signal is difficult to reproduce in a hydrostatic atmosphere, we adopted maximally strong temperature inversions, in which the temperature rises abruptly from 1000 to 3500 K at a prescribed inversion pressure. The resulting CO cross-correlation strength as a function of inversion pressure is shown in the left panel of Figure 3 (black curve), reaching a maximum signal-to-noise ratio of $\sim 7$ for an inversion near $3 \times 10^{-3}$ bar. The right panel of Figure 3 shows the corresponding dependence of the planet-to-star contrast on inversion pressure. For inversions at higher pressure, the spectrum saturates and approaches a near-blackbody at 3500 K; for inversions at lower pressure, the decreasing CO opacity progressively weakens the CO emission lines. Assuming $H_2$ as a spectrally inactive background gas with 1% CO shifts the predicted cross-correlation signal by roughly a factor of three toward higher pressures, owing to the combined effects of the reduced CO partial pressure and the increased atmospheric scale height caused by the lower mean molecular weight. Introducing $CO_2$ into the atmosphere substantially suppresses the maximum expected CO signal, as $CO_2$ efficiently blankets the CO opacity. Even a $CO_2$ abundance of only 0.1% relative to CO lowers the maximum expected cross-correlation signal-to-noise ratio to $\sim 3$.

The CO cross-correlation signal observed during Visit 1 is challenging to reconcile with a hydrostatic atmosphere unless a strong thermal inversion is confined to a narrow pressure range. This may also provide a natural explanation for the variability in the CO emission, with only marginal (~3σ) detections in Visits 2 and 4, and no signal in Visits 3 and 5 (see Supplementary Figure 1). This variability could be driven by changes in the temperature structure, for example linked to time-dependent outgassing[15] or dynamical processes within the magma ocean[16]. It may also reflect compositional changes driven by an outgassing–cloud feedback cycle, in which material released from the magma ocean forms clouds that cool the surface and temporarily suppress further outgassing; as the clouds dissipate, out-gassing resumes, producing quasi-periodic variations in atmospheric composition[17,18]. For example, a change in inversion pressure of 1–2 orders of magnitude is sufficient to reproduce the observed inter-epoch variability in CO signal strength. An optically thick but variable cloud layer could define the lower boundary of the inversion at $T_{\mathrm{low}}$, suppressing emission from deeper layers and effectively mimicking an isothermal atmosphere beneath. External scenarios, including the presence of dust or circumstellar material, have also been proposed[19]. At present, the data do not allow us to distinguish between these possibilities, but they point to an atmosphere that is dynamic rather than static.

A self-consistent grid of hydrostatic atmospheric models for 55 Cnc e (see Methods) was computed using ARCiS[21,22] to test which bulk compositions and thermal structures can reproduce the observed CO signal of Visit 1. Assuming chemical equilibrium, the grid spans a wide range of elemental abundances of C, H, O, N, S, P, Si, and Ti, following the approach of Ref. 20. This exploration indicates that atmospheres with C/O ~1 best reproduce the data, naturally driving carbon into CO rather than $CO_2$.

The presence of a strong and steep thermal inversion at approximately millibar pressures is also a key requirement to match the data. Figure 4 shows of all the self-consistent models in the grid that satisfy this criterium (although none reach $\Delta T \geq 2000$ K), together with the corresponding abundances and opacities of the main species for some representative models. While both hydrogen-rich and hydrogen-poor compositions can exhibit strong inversions, hydrogen-rich atmospheres tend to yield steeper temperature gradients in this pressure range. These atmospheres are relatively abundant in PS (phosphorus monosulfide) and TiO with high optical opacities which are driving the inversions in these cases (see Methods).

Since secondary atmospheres of rocky planets are set by the composition of the interior and subsequent outgassing, the composition of their atmospheres is directly linked to their interior redox states. The preference for hydrogen-rich models, together with the steep inversions they produce, therefore suggests an interior with relatively low oxygen fugacity, consistent with outgassing from a reduced magma ocean[23,24] (below ΔIW+2; see Methods).

We also consider whether disequilibrium effects and potentially atmospheric escape could influence the atmospheric chemistry and observed CO signal. Since $CO_2$ is more efficiently dissociated[25] in the upper atmosphere than CO, its observed low abundance could be due to photochemical depletion instead of outgassing at this level. Comparison of $CO_2$ formation and vertical mixing time scales to that of photodissociation indicate that in the case of a hydrogen-rich atmosphere, photodissociation probably does not play an important role at a millibar level due to its fast formation through CO+OH → $CO_2$. In an hydrogen-poor atmosphere the relevant time scales are more similar and photodissociation could play an important role and should be considered in future modeling.

Alternatively, a hydrodynamic outflow could potentially increase the atmospheric surface area and as such the strength of the CO signal. It would diminish the need for the very steep and strong inversion as implied for hydrostatic models, but it is generally not expected that molecules can survive the intense stellar UV field. In the unshielded UV field at the orbit of 55 Cnc e, CO is expected to survive for at most tens of minutes, a factor of a few longer than $CO_2$. This estimate neglects self-shielding[26], which can be much more effective for CO than for $CO_2$ and may therefore substantially extend the CO lifetime. Survival over timescales of hours or longer would be required for an atmospheric outflow to contribute significantly to the observed CO signal.

The results presented here suggest that 55 Cnc e, as other candidate lava planets, is a very exciting target also for future study using ground-based high-resolution spectroscopy, in particular with METIS at the Extremely Large Telescope (Supplementary Figure 2). Measurements anticipated at such high S/n will allow monitoring of its variability in CO abundance, temperature structure, and possible velocity field from atmospheric outflows.

## Methods

**Data analysis**. All analyses were performed using scripts written in Interactive Data Language (IDL 8.5.1). It focused on five spectral time series targeting secondary eclipses of 55 Cnc e, conducted as part of program 2084 (PI: Brandeker) and 1952 (PI: Hu). The five observations, all observed in SUBGIRSM64 mode with F444W filter, correspond to the following dates: Nov. 18, 2022 (Visit 1), Nov. 20, 2022 (Visit 2), Nov. 23, 2022 (Visit 3), Nov. 24, 2022 (Visit 4), and Apr. 24, 2023 (Visit 5). The fully calibrated exposure products, stage 2 with additional outlier and cosmic-ray flags (_crfints.fits; 2c), were downloaded through the STScI Mikulski Archive for Space Telescope (MAST) portal[27]. Visits 1, 2, 3, and 5 (program 2084) contain 16100 exposures in 22 files and visit 4 (program 1952) 20471 exposures in 27 files. Each exposure is 0.681 seconds. Barycentric time stamps and wavelength solutions are provided by the JWST pipeline (CRDS_VER = 13.0.6).

The extraction of 1D spectra was performed by first choosing a central exposure as reference for which the 2-dimensional spectral profile in the data frame was determined by fitting Chebyshev profiles to each row in wavelength direction. This was used to determine the contribution of each detector pixel to the total flux at a wavelength step. Background emission was ignored for this very bright star. For each exposure, the flux at each wavelength was determined by adding all flux from the pixels that contribute more than 5% to the total flux at that wavelength in the 2D spectral profile determined above, and divided by the total contribution fraction of these pixels. The next step was to shift each individual spectrum to the rest wavelength of the planet. In the case of constructing the $K_p$-$V_{sys}$ diagram of Figure 2, this step is performed for each $K_p$ separately.

The barycentric timestamps of each exposure were converted to orbital phase using the ephemeris from Ref.12, adopting an orbital period of 0.73654604 d and a transit midpoint of MJD 58723.38328. We defined the eclipse duration as $T_2 - T_3 = 89$ min and $T_1 - T_4 = 95$ min, such that exposures obtained before $T_1$ and after $T_4$ were treated as out-of-eclipse measurements (star + planet), whereas those between $T_2$ and $T_3$ were treated as in-eclipse measurements (starlight only). At each wavelength, the out-of-eclipse baseline was fitted with a linear plus exponentially decaying function[3,4,28], $F(x) = a_0(1 + a_1 e^{-a_2 x} + a_3 x)$. The linear downward trend ($a_3$) has previously been attributed to an increase in the focal-plane-array housing temperature[4]. The binned spectral time series shown in Fig. 1 additionally exhibit substantial correlated noise, dominated by an achromatic common-mode signal at the 100–200 ppm level that is likely instrumental in origin. Our cross-correlation analysis is insensitive to such achromatic variations.

Defining $\langle S_{\rm ecl} \rangle$ as the average in-eclipse spectrum and $\langle S_{\rm base} \rangle$ as the average out-of-eclipse spectrum, the planet-to-star flux ratio is given by $\langle S_{\rm base} \rangle / \langle S_{\rm ecl} \rangle - 1$.

Template spectra were generated with petitRADTRANS[13] v2.3.2, adopting the temperature structures and molecular abundances described in the main text. Opacity line lists for CO and $CO_2$ were taken from HITEMP[29,30]. We assumed a surface gravity of $2.23 \times 10^3 \rm cm\ s^{-2}$, consistent with the estimated planetary mass and radius, and divided the atmosphere into 50 layers extending down to $10^{-10}$ bar. The resulting spectra were computed at a sampling of $\lambda/\Delta\lambda = 10^6$, then convolved with a Gaussian kernel to the NIRCam resolving power ($R \approx 1600$) and binned to the wavelength grid of the observations. To convert the model contrast spectrum into a planet spectrum, we multiplied the planet-to-star contrast by a normalized, high-pass-filtered BT-Settl stellar model[31] ($T_{\rm eff} = 5200$ K, $[\rm Fe/H] = 0.3$, $\log\ g = 4.4$), convolved and binned in the same way as the NIRCam data. The stellar template was additionally boxcar-smoothed to account for its smearing in the planet rest frame. Although weak stellar CO lines appear as correspondingly weak emission features in the planet-to-star ratio spectrum, their effect on the recovered cross-correlation signal is only a few per cent.

To further validate the signal, we performed a brief, fully independent analysis of the data using principal component analysis (PCA), a standard detrending technique in ground-based high-resolution spectroscopy[7]. A PCA model was fitted to the mean-subtracted flux time series. The data were then detrended by dividing each spectrum by the PCA model plus the wavelength-channel mean. After sigma clipping, the residual flux time series were binned and cross-correlated with a CO template similar to that used in the main analysis, and the resulting cross-correlation functions were used to construct a $K_p$–$V_{sys}$ diagram. Within the uncertainties, this independent analysis recovered a signal with the same strength and location as in the fiducial reduction.

**Self-consistent atmospheric modelling**. A self-consistent grid of hydrostatic atmospheric models for 55 Cnc e was computed using ARCiS[21,22] to test which bulk compositions and thermal structures can reproduce the observed CO signal of Visit 1. The elemental composition was probed over a wide range as detailed in Supplementary Table 1, including the C/O, H/N, S/P, Si/Ti ratios, and the C+O, H+N, S+P, and Si+Ti abundances, with those of Fe, Na, K, He, Mg, Al, and Ca fixed at solar values. This grid contains 8136 distinct compositional models, covering both hydrogen-rich and hydrogen-poor cases and producing atmospheres dominated by $H_2$, CO, $CO_2$, $SO_2$, or $N_2$. The temperature structure is computed over pressures from 10 to $10^{-8}$ bar. The incident stellar spectrum is represented by a Kurucz model atmosphere for 55 Cnc A. Radiative transfer is solved using correlated-k opacity tables generated for each atmospheric composition. Molecular opacities are taken from the ExoMol database[32]. Opacities for Fe, P, Si, Mg, Ca, Al, and Ti were computed with pyROX[33] at R=10000 over 0.3–50 $\mu$m and across the relevant temperature–pressure range. Below $10^{-5}$ bar, opacities are extrapolated using values at $10^{-5}$ bar. Kurucz line lists[34] are adopted for these species. Supplementary Table 1 lists the 63 species that were included in the radiative transfer calculations. In addition, opacities for 12 collision-induced absorption (CIA) pairs were incorporated.

Chemical abundances were computed at each pressure layer using the equilibrium chemistry code GGchem[35], which solves the law of mass action through Gibbs free energy minimisation. ARCiS iterates between the chemistry and radiative transfer to reach a converged temperature structure and atmospheric composition. Heat transport within the atmosphere is considered and parameterized by the factor f, which controls the distribution of absorbed stellar flux. We adopt f = 1/4, corresponding to efficient global redistribution and nearly uniform day–night temperatures for each compositional model. Scattering is treated using a semi-3D approach in which stellar irradiation is applied only to the dayside hemisphere. Scattering therefore originates exclusively from the dayside, while the nightside remains unilluminated, preserving consistency with the assumed redistribution regime.

**Best-fit models and interpretation**. We restrict the comparison to models with CO mixing ratios greater than $10^{-2}$, and identify those satisfying the observational requirement $\mathrm{CO}/\mathrm{CO}_2 > 1000$. Models consistent with the data cluster at C/O ~ 1, producing the highest incidence of strong inversions between $10^{-2}$ and $10^{-3}$ bar.

Both hydrogen-poor and hydrogen-rich compositions can reach the required CO abundance and $CO/CO_2$ ratio. However, the steepest thermal inversions are preferentially produced in hydrogen-rich models. In these atmospheres, short-wavelength absorbers deposit stellar energy at pressures of $10^{-2}$–$10^{-3}$ bar, while longer-wavelength opacity from species such as $H_2S$ and $H_2O$ heats deeper layers. This combination produces a warm lower atmosphere, cooling toward $10^{-2}$bar, and rapid reheating at lower pressures, yielding the sharp temperature gradients required by the observed CO emission.

The preference for CO-rich, $CO_2$-poor hydrogen-rich models provides a possible link between the observed atmosphere and the redox state of the interior. For highly irradiated lava planets, the atmosphere is expected to outgass from the interior and exchange volatiles with the molten surface, so its composition may reflect the oxygen fugacity and volatile inventory of the magma reservoir[23]. Reduced or mildly reduced magma-ocean conditions naturally favour CO over $CO_2$, whereas more oxidized interiors tend to produce $CO_2$- and $SO_2$-rich atmospheres. Our preferred solutions are therefore consistent with secondary outgassing from a reduced or mildly reduced interior, although interior models remain non-unique[36] and the observed atmosphere may primarily sample the dayside magma reservoir rather than the bulk planet[19,37,38].

Opacity-removal tests show that the inversions are mainly driven by PS and TiO. Both species absorb strongly at optical wavelengths and deposit stellar energy near $10^{-2}$ – $10^{-3}$ bar, where the CO emission is most sensitive to the temperature gradient. PS is typically the dominant contributor because of its high abundance, although TiO can also sustain inversions in some oxidized cases. Their presence therefore provides a natural mechanism for producing the steep inversions required by the CO signal and offers a potential diagnostic for future observations.

**Disequilibrium effects and atmospheric escape**. Our models assume chemical equilibrium, but photochemistry and escape may alter the composition of 55 Cnc e's strongly irradiated atmosphere. We therefore compared $CO_2$ photodissociation[39], chemical formation and vertical mixing timescales for representative hydrogen-rich and hydrogen-poor models. The $CO_2$ photodissociation rate was calculated by integrating published wavelength-dependent cross sections[25] over the attenuated stellar photon flux, with attenuation treated through $CO_2$ absorption. We compared the resulting $\tau_{\mathrm{phot}} = 1/J_{\mathrm{CO_2}}$with $CO_2$ formation[40,41,42] through CO + OH → $CO_2$ + H and CO + O + M → $CO_2$ + M, and with a vertical mixing timescale $\tau_{\mathrm{mix}} = H^2/K_{zz}$, adopting $K_{zz} = 10^8\ \mathrm{cm}^2\ \mathrm{s}^{-1}$.

In hydrogen-rich atmospheres, $CO_2$ formation is faster than photodissociation throughout the modeled pressure range, implying that the low $CO_2$ abundances are not primarily caused by photochemical depletion but reflect the intrinsic composition. In hydrogen-poor atmospheres, the formation and destruction timescales can become comparable, so disequilibrium chemistry may affect the $CO_2$ abundance. However, even an oxidized model initialized with high $CO_2$ retains too much $CO_2$ to satisfy the observational constraint, indicating that photochemistry alone does not readily produce the required CO-rich, $CO_2$-poor atmosphere.

The CO signal could alternatively include a contribution from escaping gas. Highly irradiated rocky planets may drive hydrodynamic outflows[23,43], or comet-like tails[43-45], but molecular survival in such regions is limited by rapid photodissociation and photoionization[46]. CO is comparatively stable, yet maintaining large abundances in the low-density exosphere may require shielding and continuous replenishment. Although the observations are most naturally interpreted as probing denser, bound atmospheric layers, they maybe allow for a contribution from a variable outflow.

**Clouds and variability**. The CO signal is detected strongly in only one of the five eclipses, indicating substantial epoch-to-epoch variability. This behaviour is consistent with previous optical, Spitzer and JWST observations showing variable dayside emission[3,47,48] from 55 Cnc e, and suggests that the molecular signal may be intermittent rather than a persistent feature of a static atmosphere.

Several mechanisms could produce such variability. A coupled outgassing–cloud feedback cycle may modulate the surface temperature, atmospheric opacity and volatile supply: material released from the magma ocean forms clouds, which cool the surface and temporarily suppress further outgassing until the clouds dissipate[17,18]. Alternatively, the atmosphere may be transient, maintained by stochastic outgassing and partially removed by escape[15], so that molecular species are detectable only during episodes of enhanced replenishment or favourable thermal structure. Variability in the magma ocean itself, for example through tidally driven flows[16], could also alter local surface temperatures and outgassing rates.

External scenarios involving dust or circumstellar material remain possible[19], but must reproduce the observed timescales and spectral behaviour. The current data do not distinguish among these mechanisms. They do, however, indicate that 55 Cnc e is unlikely to host a single steady-state atmosphere; instead, the observations may be probing a time-variable surface–atmosphere system.

**METIS simulations**. The results presented here suggest that 55 Cnc e is also a promising target for ground-based high-resolution spectroscopy. The host star is nearby and bright, such that the stellar flux dominates over the thermal background, and the 4.6 μm CO band is comparatively free of strong telluric absorption. We simulated a single spectral time series with METIS, the Mid-infrared ELT Imager and Spectrograph, a first-generation instrument for the 39 m Extremely Large Telescope, for which first light is currently expected around 2030. The simulations are shown in Supplementary Figure 2. The continuum signal-to-noise ratio per pixel for a METIS observation of the 55 Cnc system was estimated with ScopeSim[49,50] to be 600. This noise level was applied to a spectral time series consisting of 275 exposures of 60 s each, corresponding to a total observing time of 4.6 h and covering orbital phases 0.35–0.60, similar to the JWST/NIRCam visits. The simulated spectra span 4.6–4.8 μm at a resolving power of R = 100,000, assuming two-pixel sampling.

The planetary emission spectrum was generated using the same radiative-transfer setup as used for the cross-correlation templates described in the main text. To represent a typical planetary spectrum, we adopted a step-function thermal inversion from 1000 to 3500 K at $10^{-5}$ bar. This model would yield a $\sim 3\sigma$ detection in a single JWST/NIRCam visit, comparable to the average signal strength across the five observed visits. For each exposure, the planet spectrum was Doppler-shifted into the observer frame according to the changing radial component of the planet's orbital velocity and added to a BT-Settl stellar model[31] with T_eff = 5200 K, [Fe/H] = 0.3, and log g = 4.4. Each individual exposure was then multiplied by a standard telluric transmission spectrum generated with the ESO Sky Model Calculator[51]. To remove time-stationary spectral structure, each wavelength channel was normalized by its mean value over the time series. The resulting spectra were subsequently cross-correlated with the input planet template to produce Supplementary Figure 2. The planetary signal is recovered at S/N > 20, nearly an order of magnitude stronger than the average signal obtained from the five JWST/NIRCam visits.

The high resolving power of METIS, more than a factor of 50 greater than that of NIRCam, is critical for this gain. Over the 4.6 h observing window, the changing planet radial velocity shifts the planetary spectrum by more than 200 detector pixels. This large Doppler drift makes time-stationary instrumental, telluric, and stellar features highly separable from the planet signal, allowing their removal with minimal attenuation of the planetary lines. However, these simulations should be regarded as optimistic: they include only Poisson noise and assume perfect telluric correction, perfect removal of instrumental and stellar systematics, and an exact match between the observed planet spectrum and the cross-correlation template.

**Data availability**

The data used in this paper are associated with JWST guest observer programmes 2084 and 1952 and are available from the Mikulski Archive for Space Telescopes (https://mast.stsci.edu).

**Code availability**

The radiative transfer code to generate atmospheric models is available at https://petitradtrans.readthedocs.io/en/2.7.7/ (petitRADTRANS)
Thermo-chemical equilibrium code: https://github.com/pw31/GGchem (GGchem)
Climate modelling tool: https://github.com/michielmin/ARCiS (ARCiS)
Ourgassing model: https://github.com/cvbuchem/LavAtmos (LavAtmos)
Computation of opacity cross-sections and collision-induced absorption coefficients: https://github.com/samderegt/pyROX (pyROX)

**Acknowledgements**

We thank the teams who have originally requested and executed the JWST/NIRCam observations of 55 Cnc e. This research is based on observations with the NASA/ESA/CSA JWST obtained from the Mikulski Archive for Space Telescopes at the Space Telescope Science Institute (STScI), which is operated by the Association of Universities for Research in Astronomy, Incorporated, under NASA contract NAS 5-03127. Support for this work was provided by NL-NWO Spinoza SPI.2022.004. S.d.R. and I.S. acknowledge NWO grant OCENW.M.21.010. Y.M. and L. J. acknowledge support from the European Research Council (ERC) under the European Union's Horizon 2020 research and innovation programme (grant agreement no. 101088557, N-GINE)

**Author contributions**

I.S. conceived the idea, led the data analysis, and wrote large parts of the paper. Y.M and L.J. led the physical interpretation and its write up. D.G.P. conducted the independent analysis of the data. S.d.R., N.G., and L.K. contributed to the discussions. All authors contributed to the text and figures of the paper.

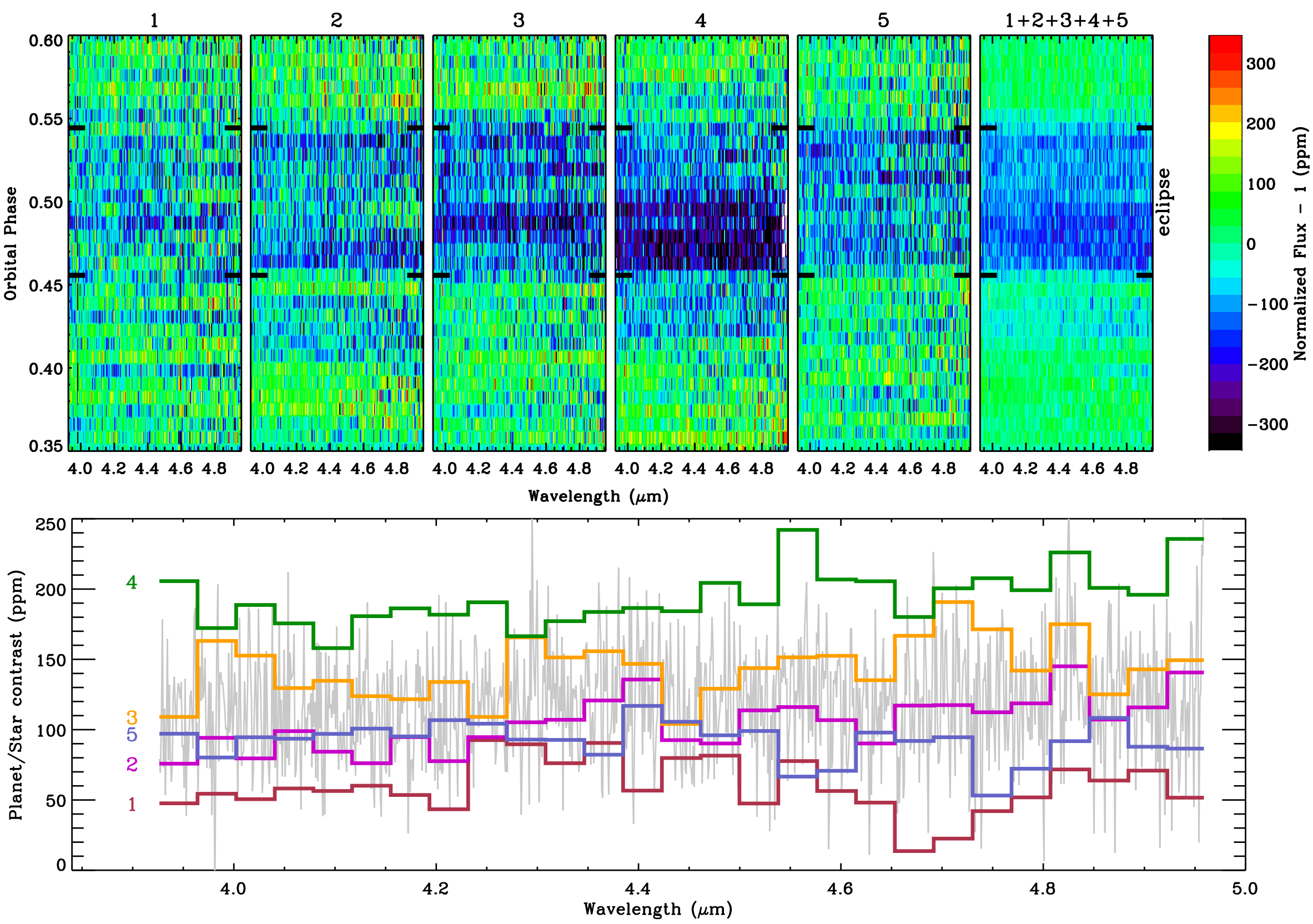

**Figure 1 | JWST/NIRCam observations of the secondary eclipse of 55 Cnc e. Top**, spectral time series obtained around secondary eclipse for Visits 1–5 and for the combined dataset. The data are binned by 5 wavelength elements and 500 time steps to reduce shot noise. The eclipse is most clearly seen in the combined time series between phases 0.46 and 0.54, with a depth of about 100 ppm. All visits are affected by achromatic common-mode noise of unknown origin, consistent with earlier studies. **Bottom**, planet-to-star contrast spectrum for each visit, binned by 50 wavelength elements. The apparent planet emission is stronger in Visit 4 (~200 ppm) than in Visit 1 (~50 ppm), although this difference may partly reflect residual common-mode noise, particularly during the first half of the eclipse in Visit 4. The average contrast spectrum at the native NIRCam spectral resolution is shown in grey.

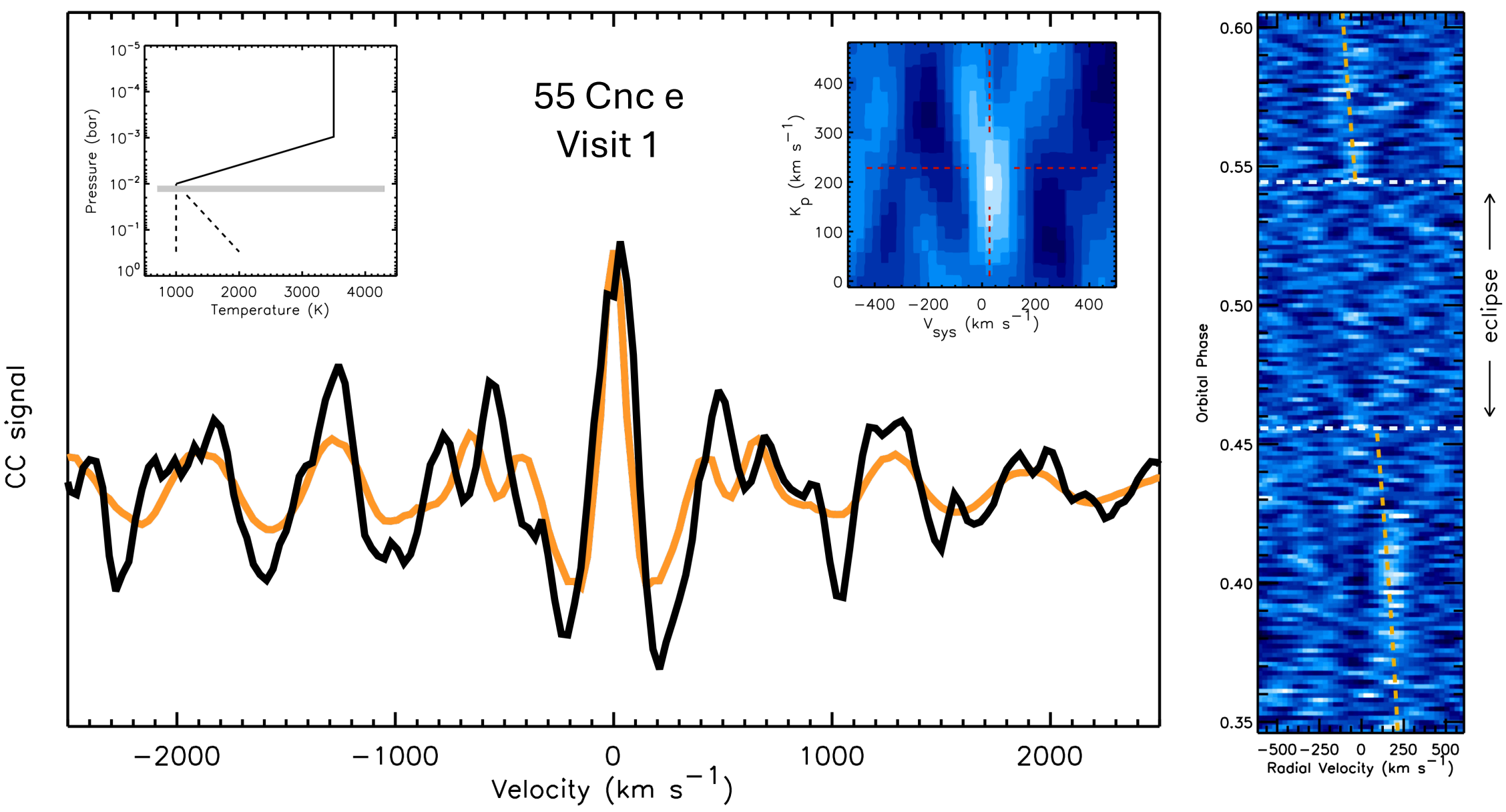

**Figure 2 | Cross-correlation detection of CO from 55 Cnc e. Left**, cross-correlation between the planet-to-star contrast spectrum in the planetary rest frame and a model template for a hydrostatic pure-CO atmosphere with an inverted temperature–pressure profile (left inset). CO is detected at a signal-to-noise ratio of ~8, with only weak dependence on the assumed atmospheric structure, including the pressure and temperature extent of the inversion and the presence of a spectrally inactive background gas such as $H_2$. An optically thick cloud deck could likewise define the base of the inversion, effectively mimicking an isothermal layer below. The inset at the right shows the $K_p - V_{sys}$ diagram[7], with the signal peaking near the expected planetary values (red dashed lines), consistent with a planetary rather than stellar origin ($K_p = 0$ km s$^{-1}$). **Right**, cross-correlation signal as a function of orbital phase shown in the barycentric rest frame. Although detected at low significance, the planetary trail follows the expected velocity pattern as indicated by the orange dashed line, appearing redshifted before eclipse and blueshifted afterwards.

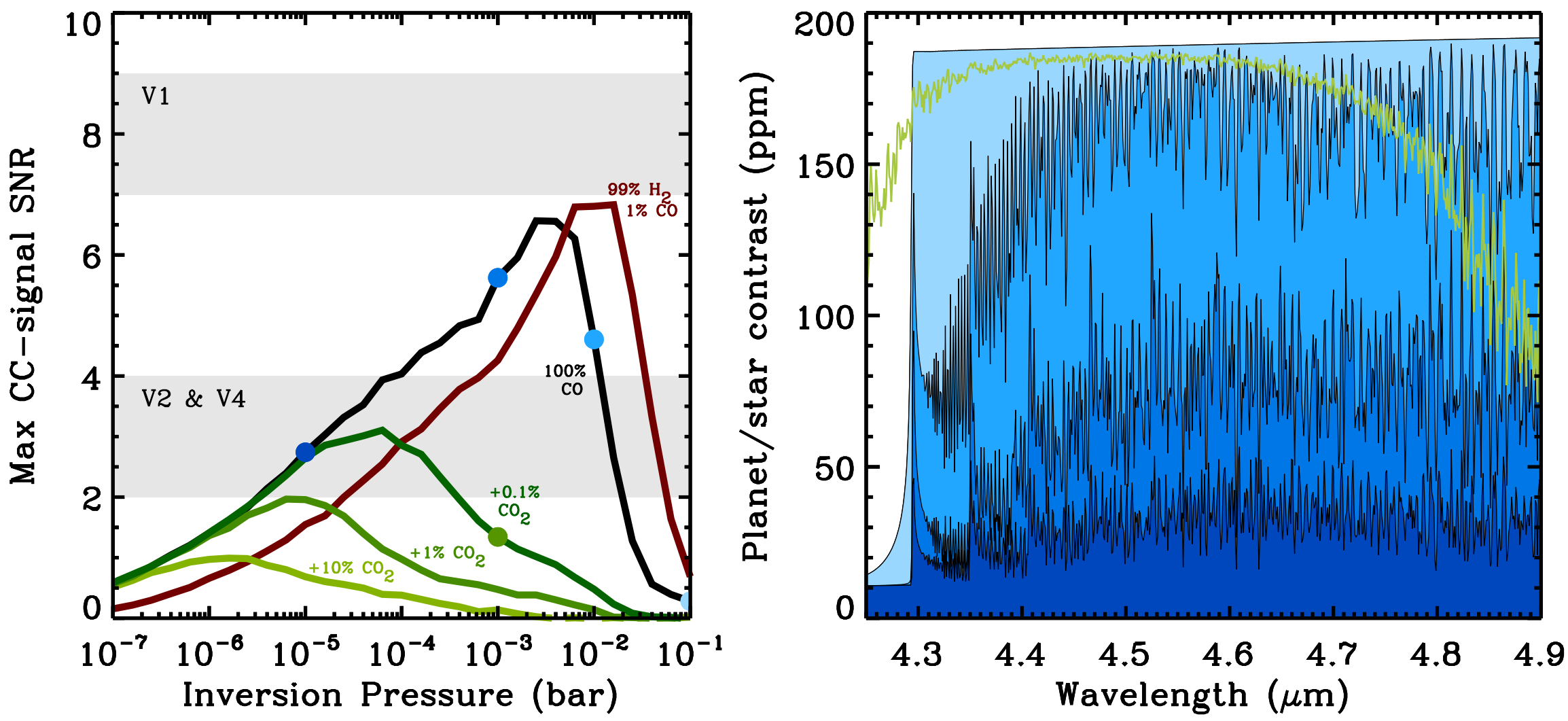

**Figure 3 | CO signal strength for synthetic models. Left panel**: Simulated cross-correlation signals, converted to signal-to-noise ratios by adding Gaussian noise at the observed level, for maximally strong step-function inversions from 1000 to 3500 K at a range of pressure levels. Results are shown for a pure CO atmosphere (black), an atmosphere with 1% CO and 99% $H_2$ (red), and atmospheres with additional $CO_2$ at 0.1%, 1%, and 10% relative to CO (green). The light-grey bands mark the observed signal-to-noise ratios for Visit 1 and Visits 2 and 4. Even for extreme thermal inversions at the optimal pressure level ($\sim 3 \times 10^{-3}$ bar), a hydrostatic atmosphere can only marginally reproduce the signal strength observed in Visit 1. **Right panel**: Spectral model templates for pure-CO atmospheres with step-function inversions at $10^{-1}$, $10^{-2}$, $10^{-3}$, and $10^{-5}$ bar, corresponding to the blue circles in the left panel. An inversion at 0.1 bar (lightest blue) nearly saturates the CO band, producing an almost featureless blackbody spectrum with little cross-correlation signal. Inversions at $10^{-2}$ to $10^{-3}$ bar yield the strongest spectral lines, whereas these are substantially weakened for an inversion at $10^{-5}$ bar. The green spectrum shows a model with an inversion pressure of $10^{-3}$ bar and $CO_2$ added at 0.1% relative to CO (green circle in the left panel). Because the opacity of $CO_2$ is several orders of magnitude greater than that of CO, even a small $CO_2$ abundance rapidly drives the spectrum toward saturation and suppresses the cross-correlation signal. The $\sim 8\sigma$ detection in Visit 1 therefore implies a remarkably high $CO/CO_2$ ratio.

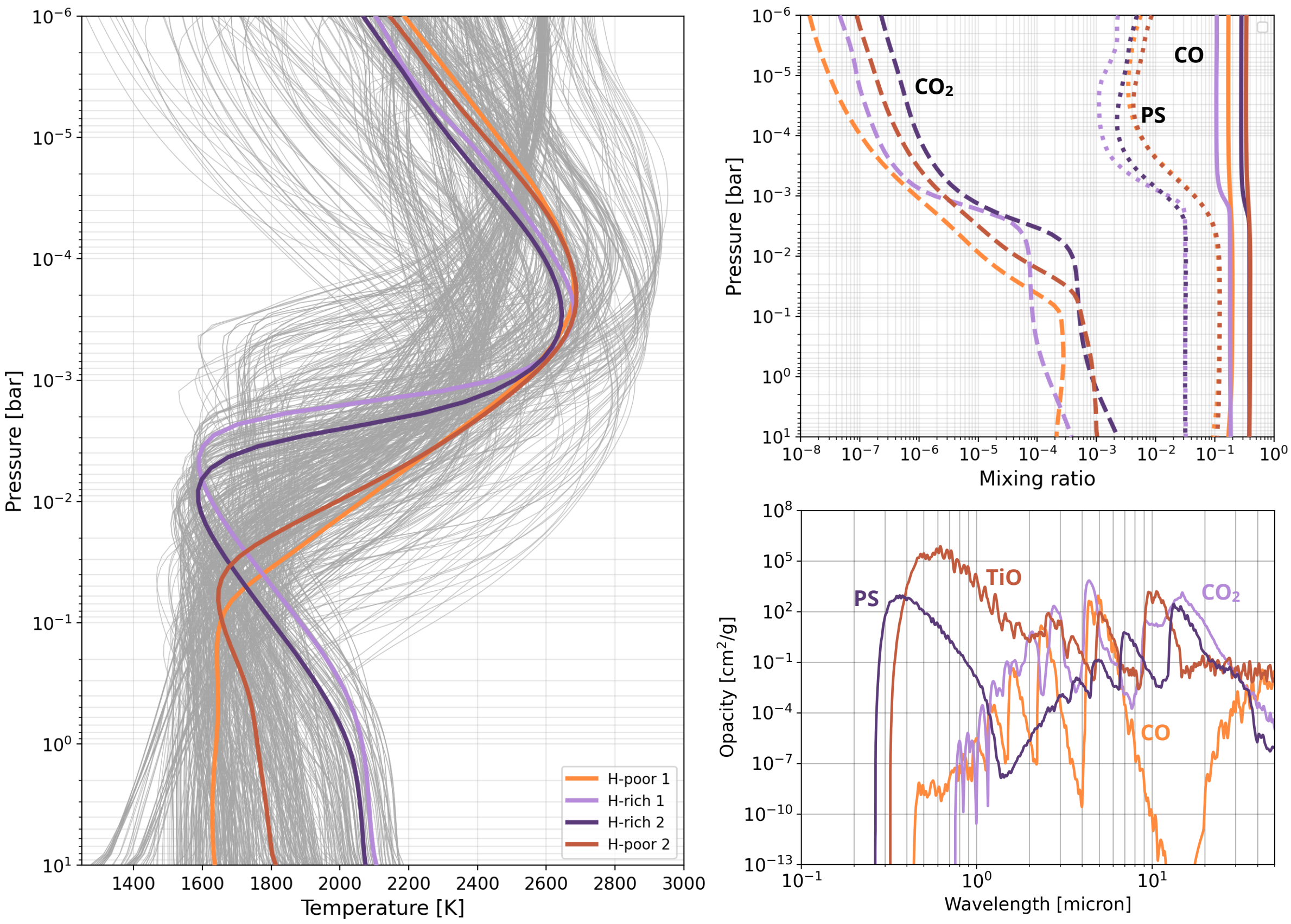

**Figure 4 | Self-consistent atmospheric models. Left,** temperature profiles for all models exhibiting strong thermal inversions near 1 to 10 mbar. Four representative models, two hydrogen-poor and two hydrogen-rich are indicated in color. **Top right**, abundances as function of pressure for CO, $CO_2$, and PS in colored solid, dashed and dotted lines respectively. **Bottom right**, opacities for PS, TiO, CO, and $CO_2$. Hydrogen-rich atmospheres in our model grid research present steeper inversions than hydrogen-poor atmospheres, driven by PS and TiO which are the primary short-wavelength absorbers. Although steep thermal inversions are readily achieved by a wide range of models in our grid, none have the extreme temperature range (> 2000 K) required to explain the CO cross-correlation signal strength of Visit 1.

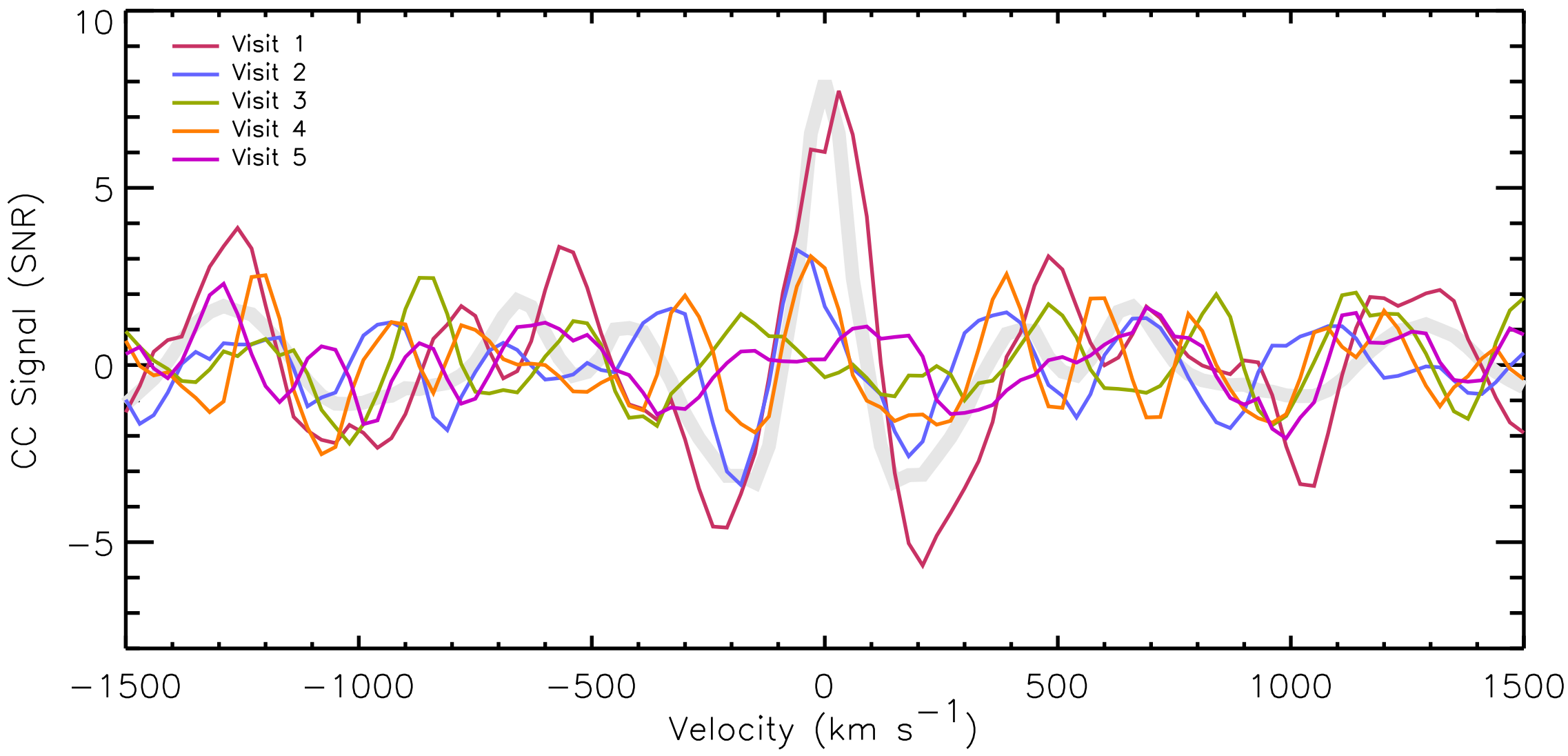

**Supplementary Figure 1 |** CO cross-correlation signals from the five individual visits, using the same spectral template as in Figure 2. The light-grey line shows the autocorrelation function, scaled to the detection significance of Visit 1. Marginal detections at the $\sim 3\sigma$ level are seen in Visits 2 and 4, while no significant signal is detected in Visits 3 and 5. This implies that the strength of the CO lines varies from visit to visit by at least a factor of a few.

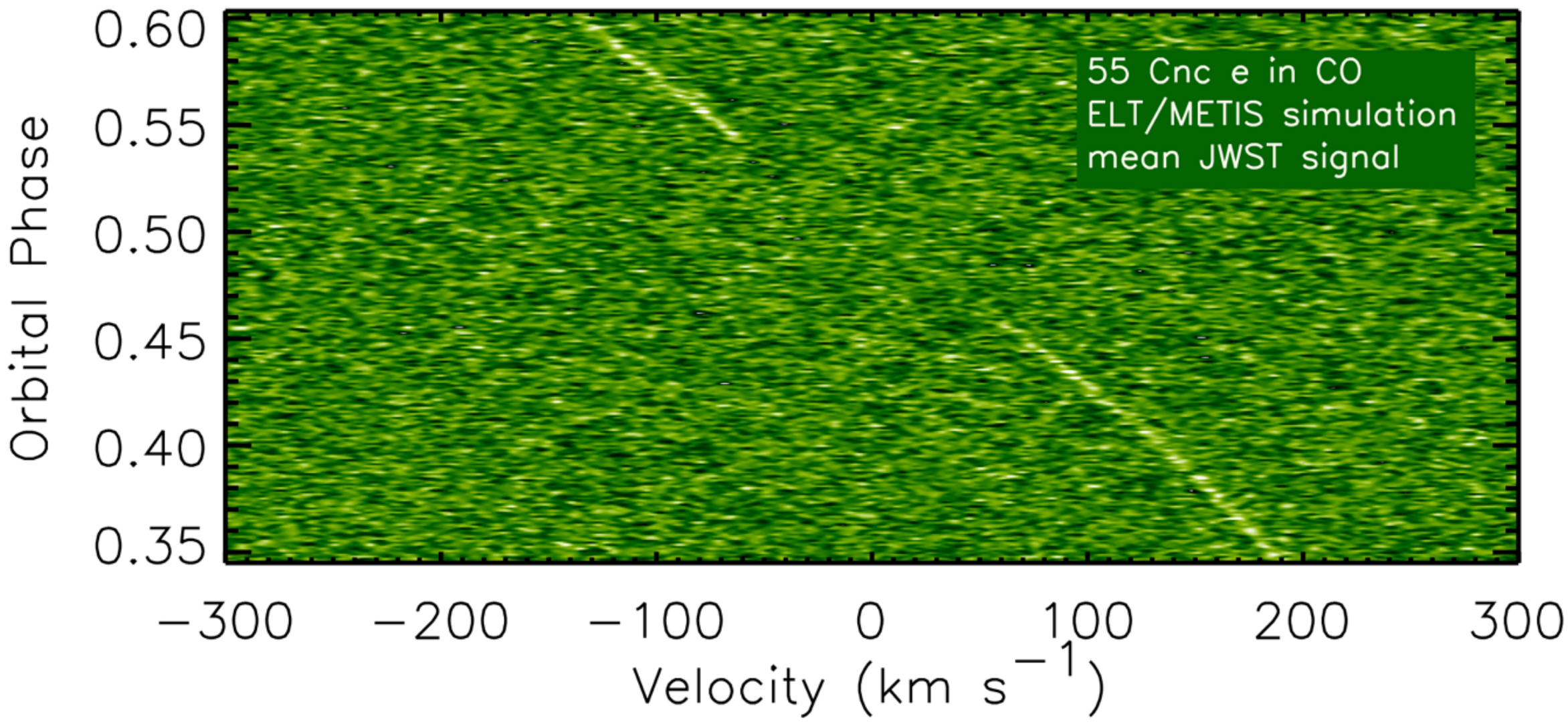

**Supplementary Figure 2 |** ELT/METIS simulations of 55 Cnc e assuming the average CO signal as detected with JWST/NIRCam. The planet signal is detected at an S/n > 20, almost an order of magnitude higher than the detection averaged over the five JWST observations. However, these simulations should be regarded as optimistic: they include only Poisson noise and assume perfect telluric correction, perfect removal of instrumental and stellar systematics, and an exact match between the observed planet spectrum and the cross-correlation template.

| Elemental Ratio | Grid range | Abundance | Grid range |
|---|---|---|---|
| C/O | $10^{-3} – 10$ | C+O | $10^{-3} – 1$ |
| H/N | $10^{-4} – 10^{8}$ | H+N | $10^{-10} – 0.4$ |
| S/P | $10^{-3} – 10^{2}$ | S+P | $10^{-20} – 0.2$ |
| Si/Ti | $10 – 10^{2}$ | Si+Ti | $10^{-20} – 0.2$ |
| Fe, Na, K, He, Mg, Al, Ca at solar values | | | |
| Species included in radiative transfer:<br>Fe, Si, Mg, Ca, Al, K, Ti, Na, N2, NO, SO, HNO3, C2, FeH, $H_2^+$, $H_3O^+$, $SiH_2$, $SiH_4$, SiS, $N_2O$, $H_2CO$, CH,CS,NH, NS,PO, $SO_3$, $SO_2$, MgH, AlH, CaH, TiH, $C_2H_2$, $C_2H_4$, SiO, AlO, CaO, $SiO_2$, $H_2O$, CO, $CH_4$, MgO,NaH, TiO, SiH, $H_2S$, PS, $CO_2$, HCN, OH, CN, HS, $NH_3$, $PH_3$, $O_2$, $H_2$, TiO, H, He, $H^-$, O, S, P. | | | |
| CIA pairs included in radiative transfer:<br>$O_2–O_2$, $CH_4–He$, $CO_2–CH_4$, $CO_2–CO_2$, $CO_2–H_2$, $CO_2–He$, $H_2–CH_4$, $N_2–H_2$, $N_2–H_2O$, $N_2–He$, $N_2–N_2$, $O_2–CO_2$ | | | |

**Supplementary Table 1 |** Overview of the grid-ranges in elemental ratios and abundances used for self-consistent modelling of the atmosphere of 55 Cnc e, and the species and CIA pairs used in the radiative transfer calculations.